\documentclass{article}
\pdfoutput=1
\usepackage{indentfirst}
\usepackage{amsmath, amsfonts}
\usepackage{bm}
\usepackage[boxruled]{algorithm2e}
\usepackage{graphicx}
\usepackage{color}
\usepackage{epstopdf}
\usepackage{geometry}
\begin{document}

\title{ Estimating Learning Effects:  A Short-Time Fourier Transform Regression Model for MEG Source Localization 
     } 
\author{Ying Yang\\ \textit{yingyan1@andrew.cmu.edu}
       \and  Michael  J. Tarr\\  \textit{michaeltarr@cmu.edu}
       \and Robert E. Kass\\ \textit{kass@stat.cmu.edu} \\
       Carnegie Mellon University \\ 
    }
\date{}

\maketitle
\textit{
This is an author manuscript. The paper is in press and will be published in the volume for the 4th NIPS Workshop on Machine Learning and Interpretation in Neuroimaging (2014),  Lecture Notes in Computer Science, Springer. (Original publication link by Springer will appear here later.)}
 
%\vspace{-2 em}
\begin{abstract}
Magnetoencephalography (MEG) has a high temporal resolution
well-suited for studying perceptual learning. However, to identify
{\it where} learning happens in the brain, one needs to apply source
localization techniques to project MEG sensor data into brain space.
Previous source localization methods, such as the short-time Fourier
transform (STFT) method by Gramfort et al.(\cite{Gramfort13}) produced
intriguing results, but they were not designed to incorporate
trial-by-trial learning effects.  Here we modify the approach in
\cite{Gramfort13} to produce an STFT-based source localization method
(STFT-R) that includes an additional regression of the STFT components
on covariates such as the behavioral learning curve.  We also exploit a
hierarchical $L_{21}$ penalty to induce structured sparsity of STFT
components and to emphasize signals from regions of interest (ROIs)
that are selected according to prior knowledge.  In reconstructing the
ROI source signals from simulated data, STFT-R achieved smaller errors
than a two-step method using the popular minimum-norm estimate (MNE),
and in a real-world human learning experiment, STFT-R yielded more
interpretable results about what time-frequency components of the ROI
signals were correlated with learning.
\end{abstract}
\section{Introduction} 
\indent Magnetoencephalography (MEG) \cite{Hamalainen93} has a high
temporal resolution well-suited for studying the neural bases of
perceptual learning.  By regressing MEG signals on covariates, for
example, trial-by-trial behavioral performance, we can identify how
neural signals change with learning.  Based on Maxwell's
equations\cite{Hamalainen94}, MEG sensor data can be approximated by a
linear transform of the underlying neural signals in a ``source
space'', often defined as $\sim10^4$ source points distributed on the
cortical surfaces.  Solving the inverse of this linear problem
(``source localization'') facilitates identifying the neural sites of
learning.  However, this inverse problem is underspecified, because
the number of sensors ($\sim 300$) is much smaller than the number of
source points.  Many source localization methods use an $L_2$ penalty
for regularization at each time point (minimum-norm estimate
\cite{Hamalainen94}, dSPM \cite{Dale00} and sLORETA
\cite{Pascual-Marqui02}).  These methods, however, may give 
noisy solutions in that they ignore the temporal smoothness of the MEG
signals.  Other methods have been proposed to capture the temporal
structure (e.g. \cite{Galka04,Lamus12}), among which, a sparse
short-time Fourier transform (STFT) method by Gramfort et
al. \cite{Gramfort13} yields solutions that are spatially sparse and
temporally smooth. \\
\indent With $L_2$ methods such as the minimum-norm estimate (MNE),
one can study learning effects in a two-step procedure: 1)~obtain
source time series in each trial; 2)~regress some
features of the time series on the covariates. However, these methods
may give noisy solutions due to lack of smoothness.  To address this,
we might want to regress the STFT components in \cite{Gramfort13} on the
covariates in a two-step procedure, but being designed for
single-trial data, \cite{Gramfort13} may not provide consistent sparse
structures across trials.  Additionally, in cases with
pre-defined regions of interest (ROIs) that are theoretically
important in perceptual learning, for example, ``face-selective'' areas
\cite{Gauthier00,Kanwisher97,Pitcher11}, it is not desirable to
shrink all source points equally to zero as in MNE.
Instead, it may be useful to assign weighted penalties to
emphasize the ROIs.\\
\indent Here we modify the model in \cite{Gramfort13} to produce a new
method (STFT-R) to estimate learning effects in MEG.  We represent
the source signals with STFT components and assume the components have
a linear relationship with the covariates.  To solve the regression
coefficients of STFT components, we design a hierarchical group lasso
($L_{21}$) penalty \cite{Jenatton11a} of three levels to induce
structured sparsity.  The first level partitions source points based
on ROIs, allowing different penalties for source points within ROIs
and outside ROIs; then for each source point, the second level encourages sparsity over
time and frequency on the regression coefficients of the STFT components, 
and finally for each STFT component, the third level induces sparsity over the coefficients for different covariates.
We derive an algorithm with an
active-set strategy to solve STFT-R, and compare STFT-R with an
alternative two-step procedure using MNE on both simulated and human
experimental data.

\section{Methods}\label{Methods}
%\subsubsection{Model}
\textbf{Model}
Assume we have $n$ sensors, $m$ source points, $T$ time points in each trial, 
and $q$ trials together. 
Let $\bm{M}^{(r)} \in  \mathbb{R}^{n \times T}$ be the sensor time series 
we observe in the $r$th trial, and $\bm{G} \in \mathbb{R}^{n \times m}$ be 
the known linear operator (``forward matrix'') 
that projects source signals to sensor space. 
Following the notation in \cite{Gramfort13}, let 
$\bm{\Phi^H} \in \mathbb{C}^{s \times T}$ 
be $s$ pre-defined STFT dictionary functions 
at different frequencies and time points (see Appendix 1). 
Suppose we have $p$ covariates (e.g. a behavioral learning curve, 
or non-parametric spline basis functions), we write them into a design matrix
 $\bm{X} \in \mathbb{R}^{q\times p}$, 
 which also includes an all-one column to represent the intercept. 
 Besides the all one column, all other columns have zero means. 
 Let the scalar $X_k^{(r)} = \bm{X} (r,k)$ be the $k$th covariate in the $r$th trial. 
When we represent the time series of the $i$th source point with STFT, 
we assume each STFT component is a linear function of the $p$ covariates:
 the $j$th STFT component in the $r$th trial is $\sum_{k=1}^p X_k^{(r)} Z_{ijk}$,
  where the regression coefficients $Z_{ijk}$'s are to be solved.  
We use a complex tensor $\bm Z\in\mathbb{C}^{ m\times s \times p}$ to denote 
the $Z_{ijk}$'s, and use $\bm{Z}_k \in \mathbb{C}^{ m\times s}$ to denote 
each layer of $\bm Z$. 
Our STFT-R model reads 
\begin{equation}
\bm{M}^{(r)} = \bm{G} \left (\sum_{k=1}^p X_k^{(r)} \bm{Z_k} \right) \bm{\Phi^H} +\bm{E}^{(r)} \hspace{1em} \text{  for } \hspace{1em} r = 1,\cdots, q. 
\nonumber
%\label{eq:model0}
\end{equation}
where the error $\bm{E}^{(r)} \in \mathbb{R}^{n \times T}$ is an i.i.d 
random matrix for each trial.
To solve $\bm{Z}$, we minimize the sum of squared prediction error across $q$ trials,  with a hierarchical
 $L_{21}$ penalty $\Omega$ on $\bm{Z}$: 
\begin{equation}
\min_{\bm{Z}}  \left(  \frac{1}{2} \sum_{r=1}^q \| \bm{M}^{(r)} - \bm{G} (\sum_{k=1}^p X_k^{(r)} \bm{Z}_k ) \bm{\Phi^H} \|_F^2 + \Omega (\bm{Z}, \alpha, \beta, \gamma,  \bm{w})  \right )\label{eq: opt_problem} 
\end{equation}
where $\|\cdot\|_F$ is the Frobenius norm and 
\begin{align}
\Omega (\bm{Z}, \alpha, \beta, \gamma, \bm{w})
 &=  \alpha \sum_l  w_l \sqrt{\sum_{i\in \mathcal{A}_l} \sum_{j=1}^s \sum_{k=1}^p |Z_{ijk}|^2  }   \label{eq: penalty_l1}\\
 &+ \beta \sum_{i=1}^m \sum_{j=1}^s \sqrt{\sum_{k=1}^p |Z_{ijk}|^2  }  \label{eq: penalty_l2}\\
 &+ \gamma \sum_{i=1}^m \sum_{j=1}^s \sum_{k=1}^p |Z_{ijk}|. \label{eq: penalty_l3} 
\end{align}
The penalty $\Omega$ involves three terms corresponding to 
three levels of nested groups, and $\alpha$, $\beta$ and $\gamma$ are tuning parameters. 
On the first level in (\ref{eq: penalty_l1}), each group under the square root 
either consists of coefficients for all source points within one ROI, 
or coefficients for one single source point outside the ROIs. 
Therefore we have $N_{\alpha}$ groups, denoted by $\mathcal{A}_l, l = 1,\cdots, N_{\alpha}$,
 where $N_{\alpha}$ is the number of ROIs plus the number of source points outside the ROIs. 
Such a structure encourages the source signals outside
the ROIs to be spatially sparse and thus reduces computational cost.
With a good choice of weights for the $N_{\alpha}$ groups,  
$\bm{w} = (w_1, w_2, \dots w_{N_{\alpha}})^T$, 
we can also make the penalty on coefficients for source points 
within the ROIs smaller than that on coefficients for source points outside the ROIs. 
On the second level, for each source point $i$, the term (\ref{eq: penalty_l2}) 
groups the $p$ regression coefficients for the $j$th STFT component under the square root, 
inducing sparsity over time points and frequencies.
Finally, on the third level, (\ref{eq: penalty_l3}) adds 
an $L_1$ penalty on each  $Z_{ijk}$ to encourage sparsity 
on the $p$ covariates, for each STFT component of each source point.\\
%
% subsection
\indent
%\subsubsection{ The FISTA algorithm }
\textbf{The FISTA algorithm} 
We use the fast iterative shrinkage-thresholding algorithm (FISTA \cite{fista}) 
 to solve (\ref{eq: opt_problem}), with a constant step size, 
 following \cite{Gramfort13}.  
Let $\bm{z}$ be a vector that is concatenated by all entries in $\bm Z$, 
and let $\bm{y}$ be a vector of the same size. 
In each FISTA step, we need the proximal operator associated 
with the hierarchical penalty $\Omega$: 
\begin{equation}
\arg \min_{ \bm z} (\frac{1}{2}\|\bm{z} - \bm{y} \|^2 + \Omega(\bm z, \alpha, \beta,  \gamma, \bm w) )
= \arg \min_{ \bm z} (\frac{1}{2}\|\bm{z} - \bm{y} \|^2 + \sum_{h=1}^N \lambda_h {\|\bm{z}|_{g_h} \|_2} )
\label{eq:prox}
\end{equation}
where we concatenate all of the nested groups on the three levels in $\Omega$ 
into an ordered list $\{ g_1, g_2, \cdots, g_N\}$ and denote the penalty on  
group $g_h$ by $\lambda_h$. 
For example,  
$\lambda_h = \alpha w_l$ if $g_h$ is the $l$th group on the first level,
$\lambda_h = \beta$ if $g_h$ is on the second level, 
and $\lambda_h = \gamma$ if $g_h$ is on in the third level.  
$\{ g_1, g_2, \cdots, g_N\}$ is obtained by listing all the third level groups, 
then the second  level and finally the first level, 
such that if $ h_1 $ is before $h_2$, 
then $ g_{h_1} \subset g_{h_2}$  or $g_{h_1} \cap g_{h_2} = \emptyset$. 
Let $\bm z|_{g_h}$ be the elements of $\bm z$  in group $g_h$. 
As proved in \cite{Jenatton11a}, (\ref{eq:prox}) is solved by 
composing the proximal operators for the $L_{21}$ penalty on each $g_h$,
following the order in the list; that is, 
initialize $\bm z \leftarrow \bm{y}$,  for $h = 1, \cdots N$ in the ordered list,
\begin{equation}
\bm{z}|_{g_h} \leftarrow  \left\{
   \begin{array}{l l} 
    \bm{z}|_{g_h} (1 - \lambda_h/\|\bm{z}|_{g_h}\|_2) & \text{ if }  \|\bm{z}|_{g_h}\|_2 > \lambda_h \\
  0  & \text{ otherwise } 
  \end{array} \right.
  \nonumber
\end{equation}   
\begin{algorithm} [h]
 \caption{FISTA algorithm given the Lipschitz constant $L$}\label{alg:FISTA}
 \KwData{ $L, f(\bm{z}) = \frac{1}{2} \sum_{r=1}^q\| \bm{M}^{(r)} - \bm{G} \left (\sum_{k=1}^p X_k^{(r)} \bm{Z_k} \right) \bm{\Phi^H} \|_F^2, \Omega(\bm{z}) = \Omega (\bm{Z}, \alpha, \beta, \gamma, \bm{w})$}
 \KwResult{ the optimal solution $\bm z$ }
 initialization: $\bm{z_0}, \zeta = 1, \zeta_0 = 1$,  $\bm{y} \leftarrow \bm{z_0}$,  $\bm{z} \leftarrow \bm{z_0}$ \;
 \While { $\text{change of $\bm z$ in two iterations is not small enough}$ }{
 $\bm{z_0} \leftarrow \bm{z}$;  Compute $\nabla f(\bm{y})$ \;
 Apply the proximal operator 
   $\bm{z} = \arg_{\bm{x}}\min(\frac{1}{2} \|\bm{x} - (\bm{y}-\frac{1}{L} \nabla f(\bm{y}))\|^2 + \frac{1}{L} \Omega(\bm{x}))$\; 
  $\zeta_0 \leftarrow \zeta$;   
  $\zeta \leftarrow\frac{1+\sqrt{4\zeta_0^2+1}}{2}$;  
  $\bm{y} \leftarrow \bm{z} + \frac{\zeta_0-1}{\zeta} (\bm{z}-\bm{z_0})$\; 
}
\end{algorithm}
Details of FISTA are shown in Algorithm~\ref{alg:FISTA}, 
where $\bm{y}$ and $\bm{z_0}$ are auxiliary variables of the same shape as $\bm{z}$, 
and $\zeta, \zeta_0$ are constants used to accelerate convergence. 
The gradient of $f(\bm{z})$ is computed in the following way:
$
\frac{\partial f}{\partial \bm{Z}_k}  
 = -\bm{G}^T \sum_{r=1}^q  X_{k}^{(r)} \bm{M}^{(r)} \bm{\Phi} 
  + \bm{G}^T \bm{G} (\sum_{r=1}^q  X_{k}^{(r)} \sum_{k'=1}^p  \bm{Z}_{k'} X_{k'}(r))
  \bm{\Phi}^{H} \bm{\Phi}.
$
We use the power iteration method in \cite{Gramfort13} to 
compute the Lipschitz constant of the gradient. \\
% subsection
%
\indent 
%\subsubsection{The active-set strategy}
\textbf{The active-set strategy}
In practice, it is expensive to solve the original problem in (\ref{eq: opt_problem}).  
Thus we derive an active-set strategy (Algorithm~\ref{alg:active_set}), according to Chapter 6 in \cite{Bach11}: 
starting with a union of some groups on the first level 
($J = \cup_{l \in \mathcal{B}}{ \mathcal{A}_l }, \mathcal{B} \subset \{1, \cdots, N_{\alpha}\}$), 
 we compute the solution to the problem constrained on $J$,
  then examine whether it is optimal for the original problem 
  by checking whether the Karush-Kuhn-Tucker(KKT) conditions are met, 
  if yes, we accept it, otherwise, we greedily add more groups to $J$ and 
  repeat the procedure.\\
\indent
Let $\bm z$ denote the concatenated $\bm Z$ again, 
and let diagonal matrix $\bm{D}_h$ be a filter to 
select the elements of $\bm z$ in group $g_h$ 
(i.e. entries of $ \bm{D}_h \bm z$ in group $g_h$ are equal to $\bm z|_{g_h}$,
and entries outside $g_h$ are 0).
Given a solution $\bm{z}_0$, the KKT conditions are 
\begin{equation}
\nabla f(\bm{z})_{\bm z = \bm{z}_0} + \sum_h \bm{D}_h \bm{\xi}_h = 0, \text{and}
\begin{cases}
    \bm{\xi}_h =  \lambda_h \frac{ \bm{D}_h \bm{z}_0 } {\| \bm{D}_h \bm{z}_0 \|_2} &\text{if} \, \| \bm{D}_h \bm{z}_0 \|_2 > 0, 
    \\
    \|\bm{\xi}_h\|_2  \le \lambda_h  
    &\text{if} \, \| \bm{D}_h \bm{z}_0 \|_2 = 0
\end{cases}
\nonumber
\end{equation}
where $\bm {\xi}_h, h = 1, \cdots, N$ are Lagrange multipliers 
of the same shape as $\bm z$.  We defer the derivations to Appendix 2.\\
\indent 
We minimize the following problem 
\begin{align*}
&\min_{\bm{\xi}_h, \forall h}  \frac{1}{2} \|\nabla f(\bm{z})_{\bm z = \bm{z}_0} + \sum_h \bm{D}_h \bm{\xi}_h\|_2^2, \\
\text{ subject to} & 
  \begin{cases}
    \bm{\xi}_h =  \lambda_h \frac{ \bm{D}_h \bm{z}_0 } {\| \bm{D}_h \bm{z}_0 \|_2} &\text{if} \, \| \bm{D}_h \bm{z}_0 \|_2 > 0, 
    \\
    \|\bm{\xi}_h\|_2  \le \lambda_h  
    &\text{if} \, \| \bm{D}_h \bm{z}_0 \|_2 = 0
    \end{cases}
\end{align*}
and use $\frac{1}{2} \|\nabla f(\bm{z})_{\bm z = \bm{z}_0} + \sum_h
\bm{D}_h \bm{\xi}_h\|_2^2$ at the optimum to measure the violation of KKT conditions. 
Additionally, we use the $\frac{1}{2} \| \left( \nabla f(\bm{z})_{\bm z = \bm{z}_0} + \sum_h
\bm{D}_h \bm{\xi}_h\right)|_{\mathcal{A}_l}\|_2^2$, constrained on each non-active first-level group $\mathcal{A}_l \not\subset J$, as a measurement of violation for the group. 
\begin{algorithm} [h] 
 \caption{Active-set strategy}\label{alg:active_set}
 initialization: choose initial $J$ and initial solution $Z$; compute the KKT violation for each $\mathcal{A}_l \not\in J$ \;
 \While{ the total KKT violation is not small enough}{
  add 50 non-active groups that have the largest KKT violations to $J$\;
  compute a solution to the problem constrained on $J$ using FISTA \;
  compute the KKT violation for each $\mathcal{A}_l \not\subset J$  \;  
}
\end{algorithm}

%\subsubsection{$L_2$ regularization and bootstrapping}
\textbf{$\bm{L_2}$ regularization and bootstrapping}
The hierarchical $L_{21}$ penalty may give biased results \cite{Gramfort13}. 
To reduce bias, we computed an $L_2$ solution constrained on the non-zero entries 
of the hierarchical $L_{21}$ solution. 
Tuning parameters in the $L_{21}$ and $L_2$ models were selected to minimize 
cross-validated prediction error. \\
\indent
To obtain the standard deviations of the regression coefficients in $\bm Z$, 
we performed a data-splitting bootstrapping procedure.
The data was split to two halves (odd and even trials).  
On the first half, we obtained the hierarchical $L_{21}$ solution, 
and on the second half, we computed an $L_2$ solution constrained 
on the non-zero entries of the hierarchical $L_{21}$ solution. 
Then we plugged in this $L_2$ solution $\bm Z$ to obtain residual sensor time series 
of each trial on the second half of the data ($\bm{R} ^{(r)} = \bm{M}^{(r)}- \bm{G}(\sum_{k=1}^p X_k^{(r)} \bm{Z}_k)\bm{\Phi^H}$). 
We rescaled the residuals according to the design matrix $\bm X$ \cite{Stine85}. 
Let $ X_r = \bm{X}(r,:)^T = (X_1^{(r)}, X_2^{(r)}, \cdots, X_p^{(r)})^T$,  and  $h_r = X_r^T (\bm{X}^T  \bm{X})^{-1} X_r$. 
The residual in the $r$th trial was rescaled by $ 1/(1-h_r)^{0.5}$. 
The re-sampled residuals $\bm{R} ^{(r)*}$s were
 random samples with replacement from $\{\bm{R} ^{(r)} /(1-h_r)^{0.5}, r = 1,\cdots,q\}$
and the bootstrapped sensor data for each trial were 
\[\bm{M}^{(r)*} = \bm{G} (\sum_{k=1}^p X_k^{(r)} \bm{Z}_k) \bm{\Phi^H} + \bm{R} ^{(r)*}\] 
After $B$ re-sampling procedures, for each bootstrapped sample, 
we re-estimated the solution to the $L_2$ problem 
constrained on the non-zero entries again, 
and the best $L_2$ parameter was determined by a 2-fold cross-validation. \\

\section{Results}
\textbf{Simulation} On simulated data, we compared STFT-R with an
alternative two-step MNE method (labelled as MNE-R), that is, (1) obtain MNE
source solutions for each trial; (2) apply STFT and regress the
STFT components on the covariates.\\
\indent
We performed simulations using ``mne-python'' \cite{mne-python},
which provided a sample dataset, and a source space that consisted 
of 7498 source points perpendicular to the gray-white matter boundary,
 following the major current directions that MEG is sensitive to. 
Simulated source signals were constrained in four regions 
in the left and right primary visual and auditory cortices 
(\texttt{Aud-lh}, \texttt{Aud-rh}, 
\texttt{Vis-lh} and \texttt{Vis-rh}, Fig.~\ref{fig:simu1}(a)). 
All source points outside the four regions were set to zero.
To test whether STFT-R could emphasize regions of interest,
 we treated \texttt{Aud-lh} and \texttt{Aud-rh} as the target ROIs 
 and \texttt{Vis-lh} and \texttt{Vis-rh} as irrelevant signal sources. 
The noiseless signals were low-frequency Gabor functions 
(Fig.~\ref{fig:simu1}(b)), whose amplitude was a linear function of a sigmoid curve 
(simulated ``behavorial learning curve'', Fig.~\ref{fig:simu1}(c)). 
We added Gaussian process noise on each source point 
in the four regions independently for each trial. 
We denoted the ratio of the marginal standard deviation of this noise to the largest 
amplitude of the signal as \textit{noise level}s, 
and ran multiple simulations with different \textit{noise level}s. 
We also simulated the sensor noise as multivariate Gaussian noise 
filtered by a 5th order infinite impulse response (IIR) filter. 
The filter and covariance matrix of the sensor noise 
were estimated from the sample data. 
We used different signal-to-noise ratios (\text{SNR}s) in Decibel 
when adding sensor noise. 
Hence we had two different levels of noise characterized by 
\textit{noise level} in the source space and \textit{SNR} in the sensor space. \\
\indent
We ran 5 independent simulations for \textit{SNR} $\in\{0.5,1\}$ and \textit{noise level} $\in\{0.1,0.3,0.5\}$,
 with 20 trials (length of time series $T = 100$, sampling rate  = 100~Hz,
  window size of the STFT  = $160$ ms and step size $\tau_0 = 40$~ms).
With only one covariate (the sigmoid curve), 
we fit an intercept and a slope for each STFT component. 
Before applying both methods, we pre-whitened the correlation between sensors. 
In STFT-R, the weights for $\alpha$ in the ROI groups were set to zero,
and the weights in the non-ROI groups were equal and summed to 1. 
We tuned the penalization parameters $\alpha$, $\beta$ in STFT-R. 
For $\gamma$, because the true slope and intercept were equal in the simulation, we did not need a large $\gamma$ to select between the slope and intercept, therefore we fixed $\gamma$ to a small value to reduce the time for parameter tuning.
The $L_2$ penalty parameter in MNE-R was also selected via cross-validation.
We used $B= 20$ in bootstrapping.\\
% set the noice level
\newcommand{\snr}{0.5}
\newcommand{\nratio}{0.5}
\indent
We reconstructed the source signals in each trial using the estimated $\bm{Z}$. 
Note that true source currents that were close to each other 
could have opposite directions due to the folding of sulci and gyri, 
and with limited precision of the forward matrix, 
the estimated signal could have an opposite sign to the truth. 
Therefore we ``rectified'' the reconstructed signals and the true 
noiseless signals by taking their absolute values, and computed the 
mean squared error (MSE) on the absolute values. 
 Fig.~\ref{fig:simu1}(d) shows estimated source signals in the target ROIs
(red and yellow) by the two methods in the $20$th trial 
(\textit{SNR} = \snr, \textit{noise level} = \nratio). 
Noticing the scales, we found that MNE-R shrank the signals much more than STFT-R. 
We show the ratios of the rectified MSEs of
STFT-R to the rectified MSEs of MNE-R for source points within the ROIs (Fig.~\ref{fig:simu1}(e)), 
and for all source points in the source space (Fig.~\ref{fig:simu1}(f)). 
Compared with MNE-R, STFT-R reduced the MSE within the ROIs by about $20\sim 40\%$ (Fig.~\ref{fig:simu1}(e)) . 
STFT-R also reduced the MSE of all the source points by about $20\%$ in cases with low \textit{noise level}s (0.1) (Fig.~\ref{fig:simu1}(f)). 
The MSE reduction was larger when \textit{noise level} was small.\\ 
\begin{figure}[htbp]
\centering
\begin{tabular}{ccc}
% l b r t
\includegraphics[trim = 13 60 30 90 mm, clip = true, width = 0.25 \textwidth]{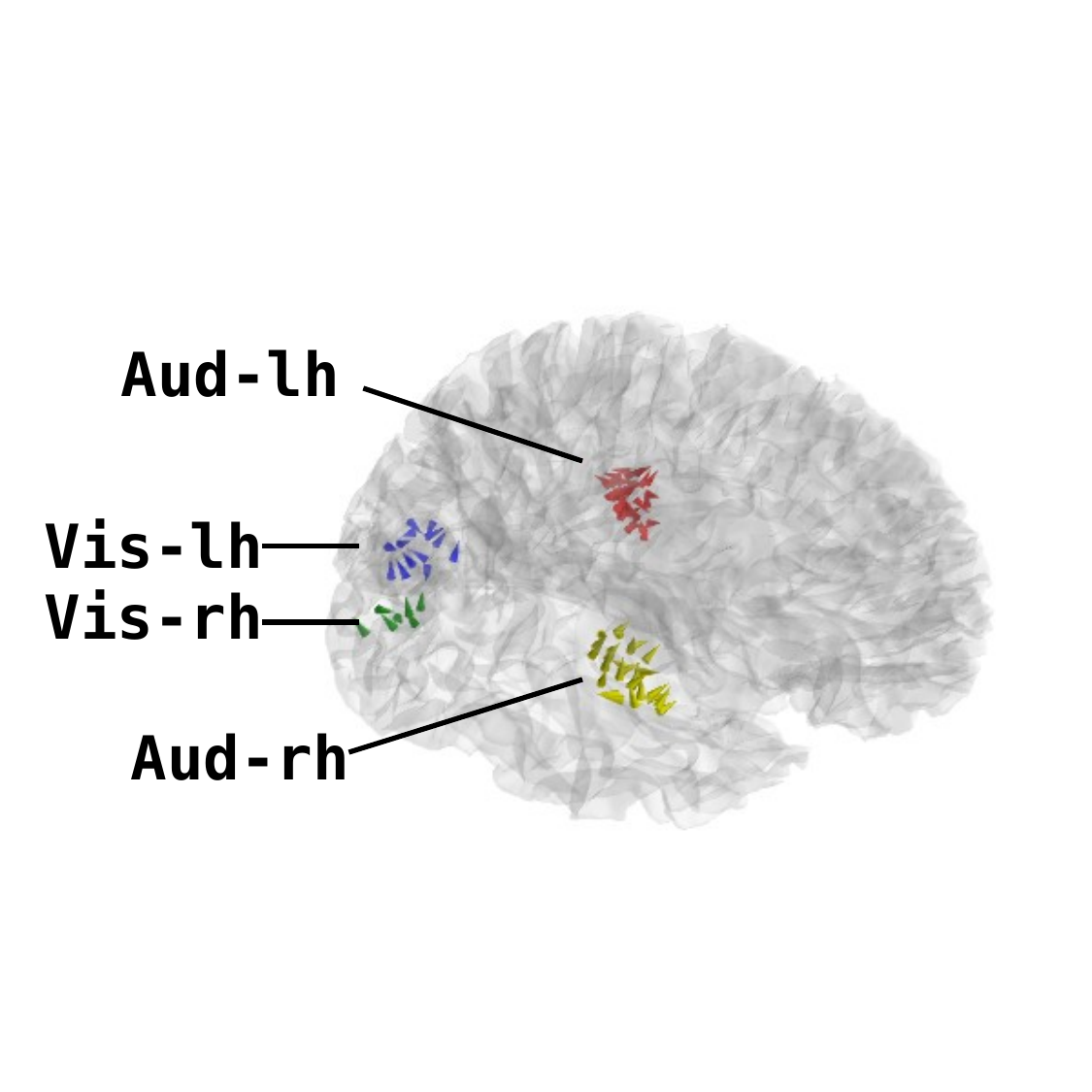}
&
\includegraphics[trim = 0 0 20 0 mm, clip = true, width = 0.33 \textwidth]
{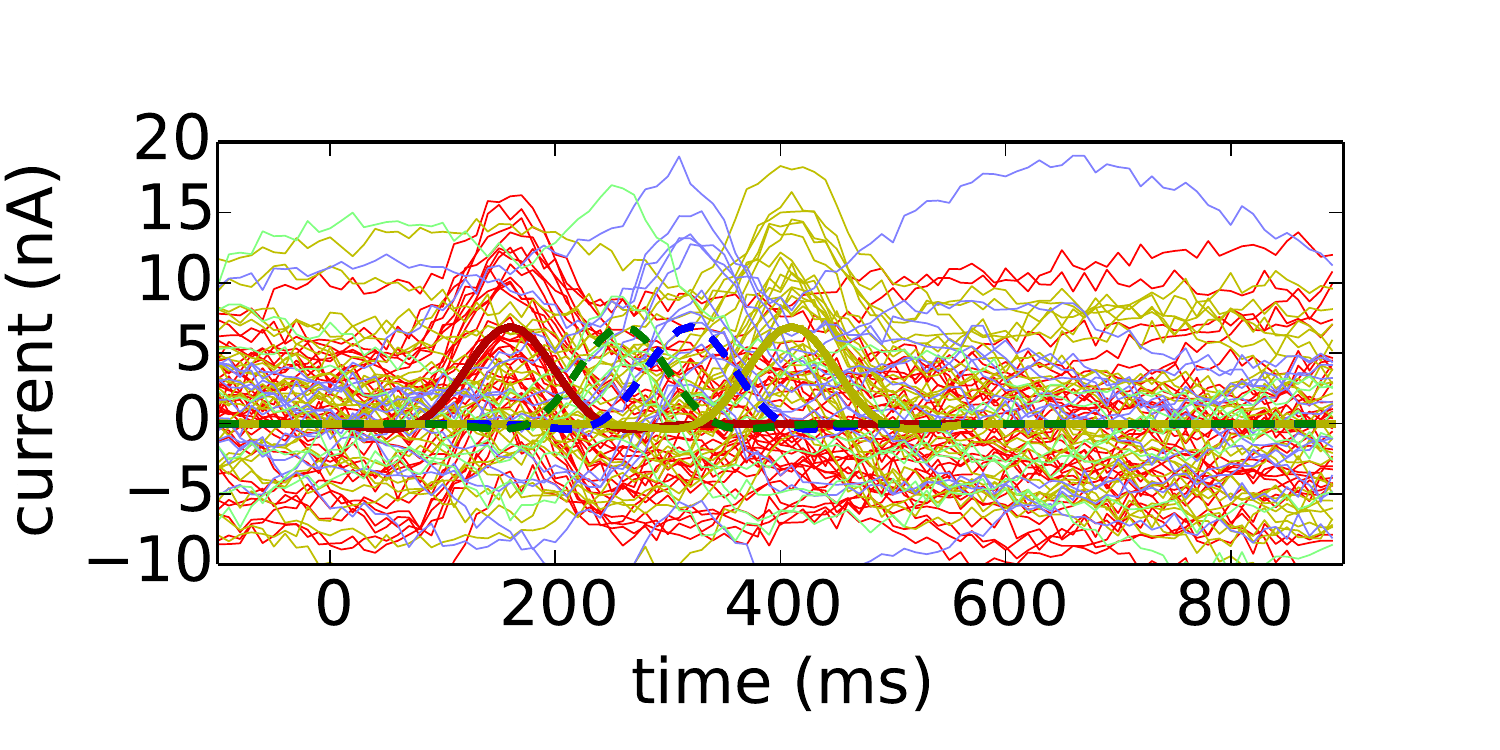}
&
\includegraphics[width = 0.32\textwidth ]{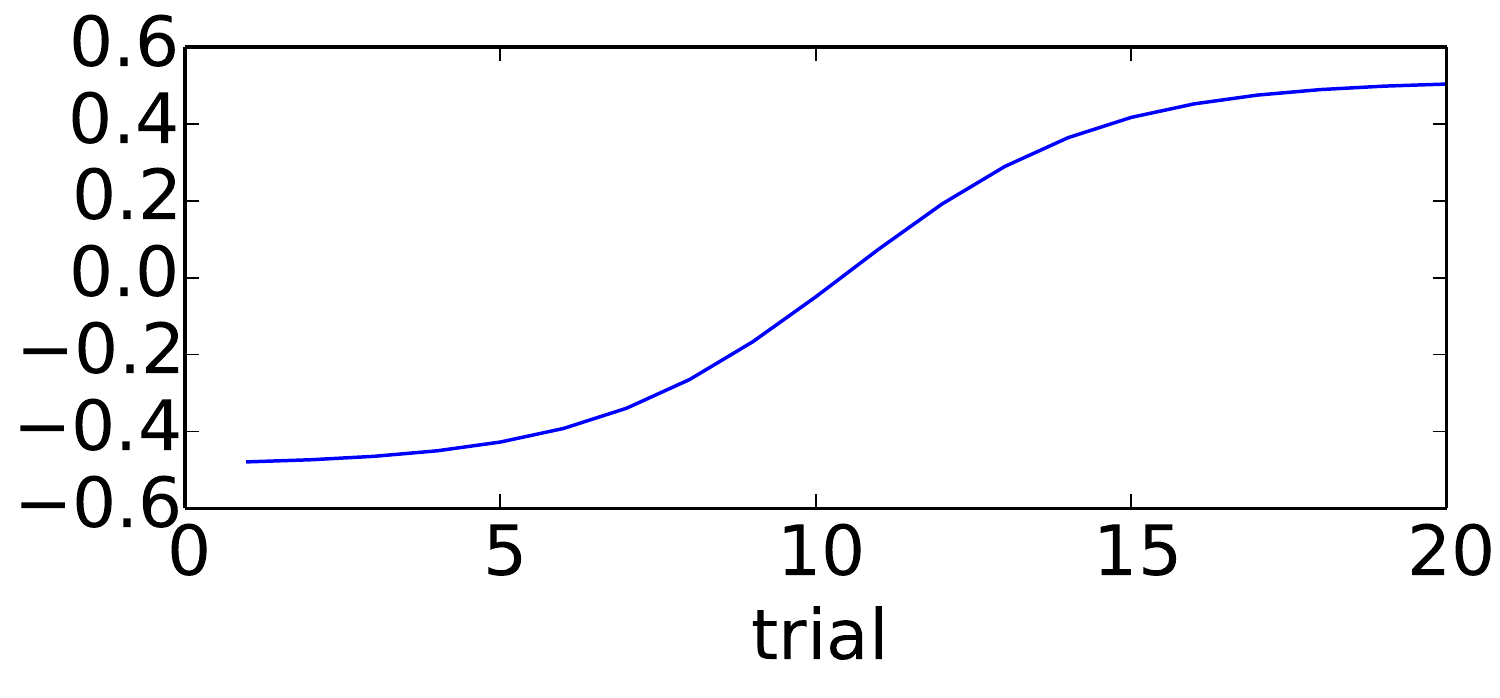} \\
(a) & (b) & (c)\\
\end{tabular}
\begin{tabular}{ccc}
\includegraphics[trim = 17 0 30 15 mm, clip = true, width = 0.34 \textwidth]{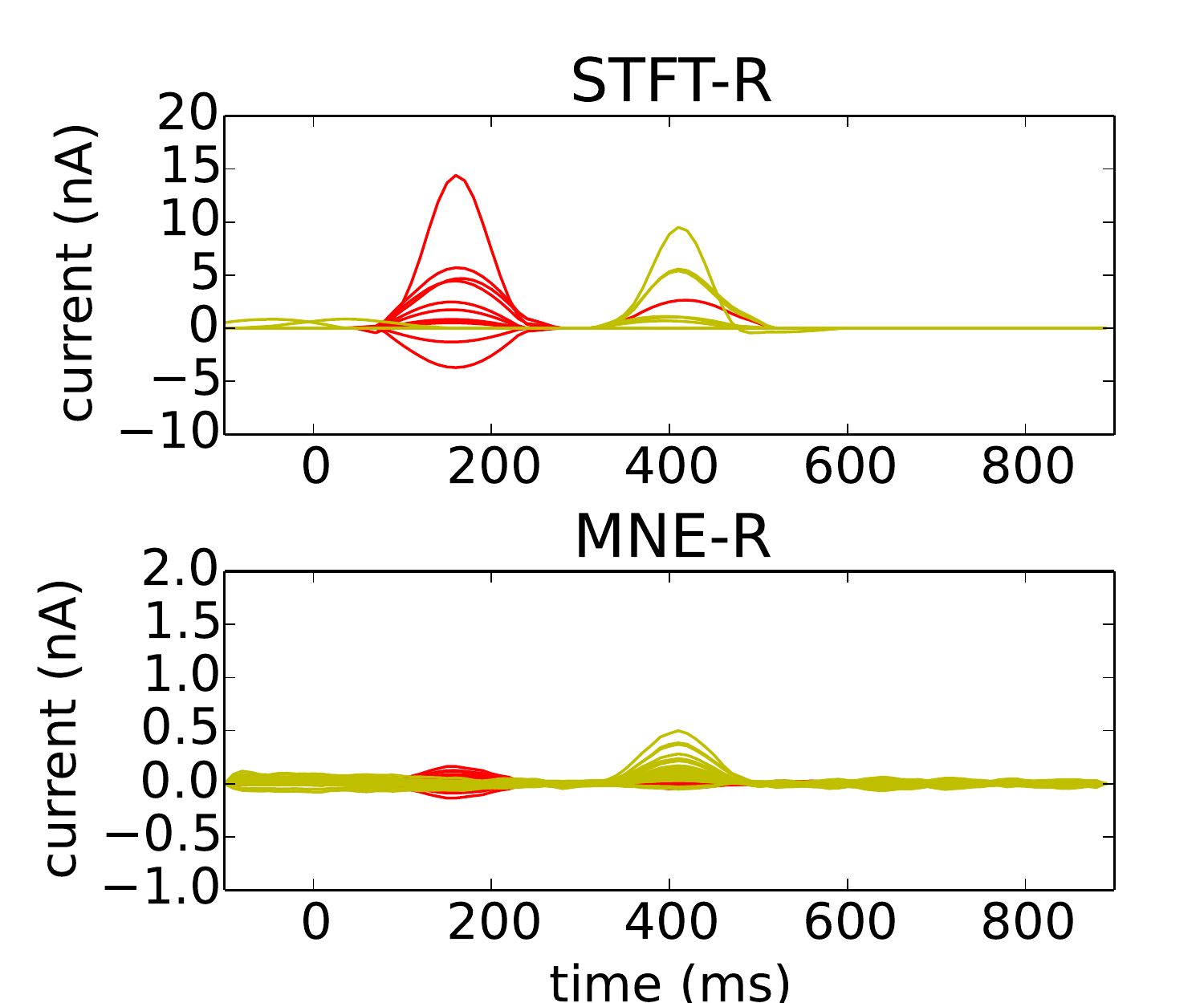} & 
\includegraphics[trim = 0 0 30 20 mm, clip = true, width = 0.2 \textwidth]{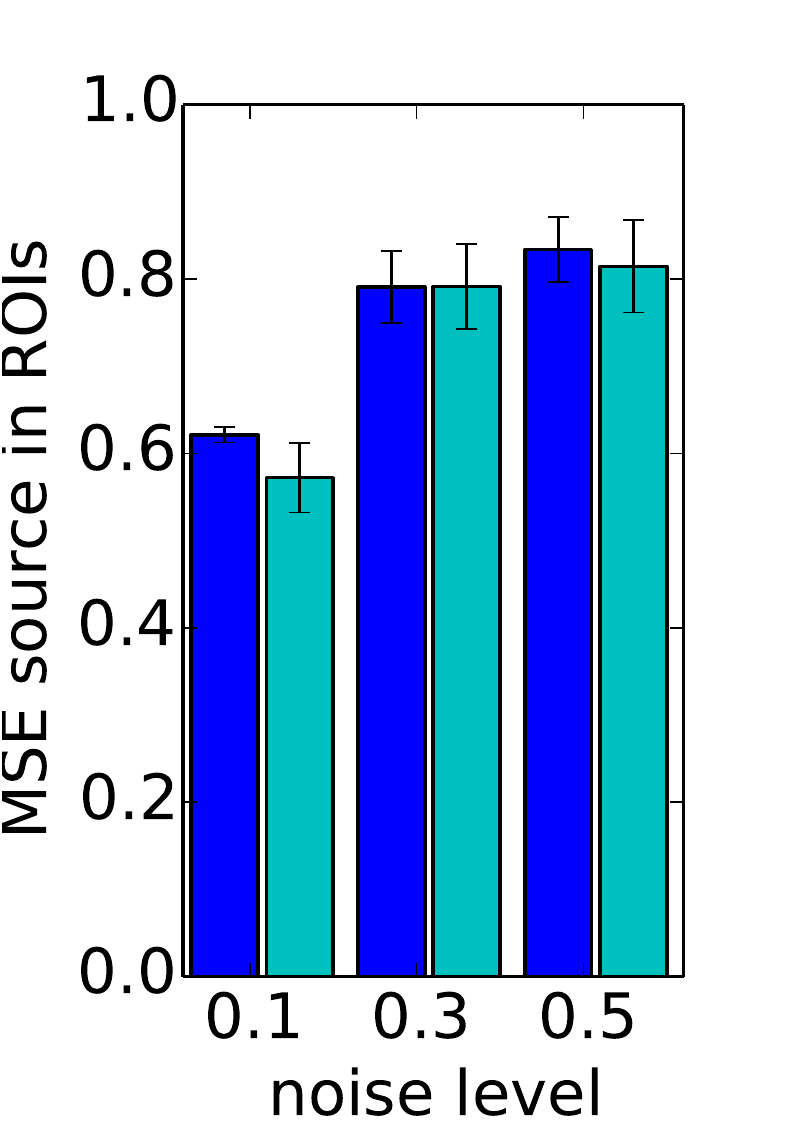} 
& \includegraphics[trim = 0 0 30 20 mm, clip = true,  width = 0.2 \textwidth]{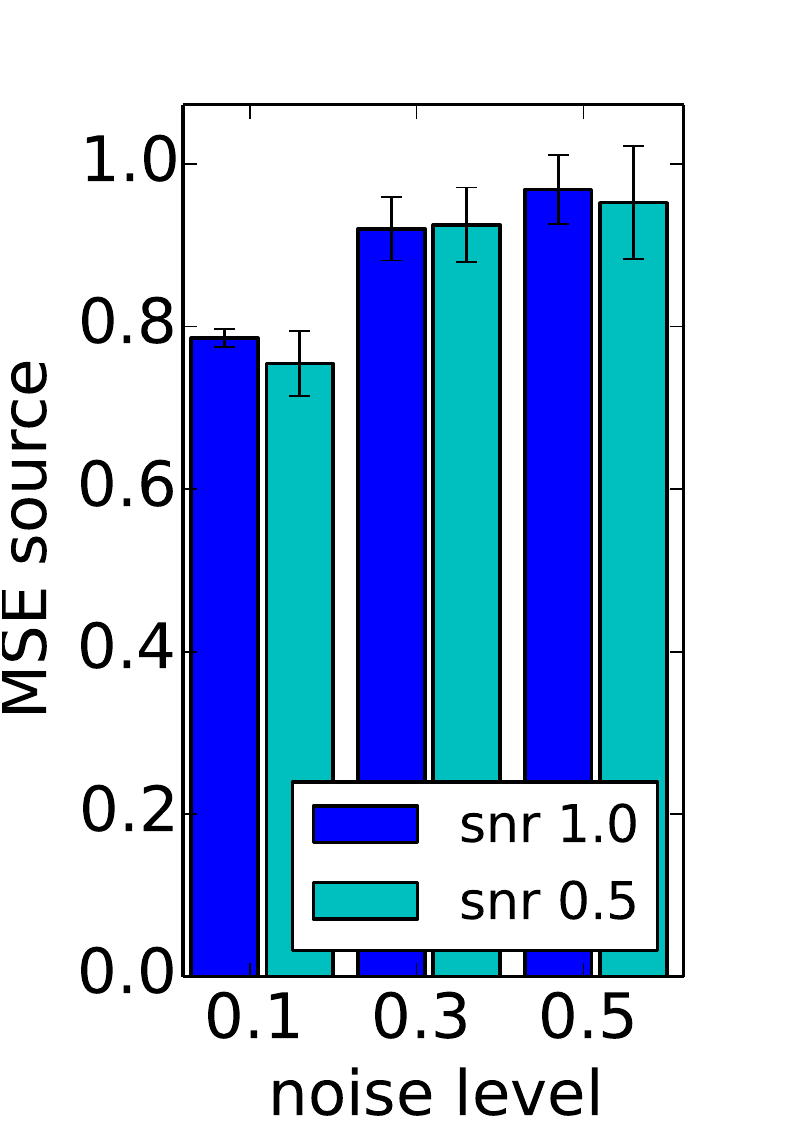}\\
(d) & (e) &(f)\\
\end{tabular}
\caption{ Simulation results: source signal reconstruction.  
(a), Target ROIs: \texttt{Aud-lh} (red), \texttt{Aud-rh} (yellow) 
 and irrelevant regions:  \texttt{Vis-lh} (blue), \texttt{Vis-rh} (green).
(b), The simulated source signals with Gaussian process noise in the 20th trial. 
Each curve represents one source point. 
The thicker curves show the noiseless true signals 
(solid: target ROIs, dashed: irrelevant regions). \textit{noise level} = \nratio.
(c), The simulated ``behavioral learning curve''. 
(d), Estimates of source signals (\textit{noise level} = \nratio, \textit{SNR} = \snr) in the $20$th trial by STFT-R and MNE-R,
in \texttt{Aud-lh} (red) and \texttt{Aud-rh} (yellow). Each curve represents one source point. 
Note the scale for MNE-R is $< 1/10$ of the truth. 
(e) and (f), Ratios of rectified MSE (STFT-R over MNE-R) for source points within the target ROIs (e) and for all source points (f). 
The bars show averaged ratios across 5 independent runs of simulation, and the error bars show standard errors of the averaged ratios. 
 }\label{fig:simu1}
\end{figure}
\indent 
To visualize which time-frequency components were correlated with the covariate, 
we computed the T-statistic for each slope coefficient of each STFT component, 
defined as the estimated coefficient divided by the bootstrapped standard error. 
Again, since our estimate could have an opposite sign to the true signals, we rectified the T-statistics by using their absolute values. 
We first averaged the absolute values of the T-statistics for the real and imaginary parts of each STFT component,
and then averaged them across all non-zero source points in an ROI, for each STFT component.
We call these values \textit{averaged absolute T}s. \\
\indent
In Fig.~\ref{fig:simu2}, we plot the T-statistic of the slope coefficient 
for each STFT component of each source point in the two ROIs 
by STFT-R (Fig.~\ref{fig:simu2}(b)) and
by MNE-R (Fig.~\ref{fig:simu2}(c)),
and compared them with the true coefficients in Fig.~\ref{fig:simu2}(a)
(\textit{SNR} = \snr, \textit{noise level} = \nratio). 
The T-statistics for the real and imaginary parts are shown separately.
In Fig.~\ref{fig:simu2}(a),(b) and (c), 
the vertical axis corresponds to the indices of source points, 
concatenated for the two ROIs. 
The horizontal axis corresponds to the indices of STFT components, 
which is a one-dimensional concatenation of the cells of 
the frequency $\times$ time matrix in Fig.~\ref{fig:simu2}(d);
0-24 are 25 time points in 0 Hz, 25-49 in 6.25 Hz, 50-74 in 12.5 Hz, and so on.
STFT-R yielded a sparse pattern, 
where only the lower frequency (0 to 6.25 Hz) components were active, 
whereas the pattern by MNE-R spread into higher frequencies 
(index 100-200,25-50Hz).
We also compared the \textit{averaged absolute T}s for each ROI by STFT-R 
(Fig.~\ref{fig:simu2}(e)) and  by MNE-R (Fig.~\ref{fig:simu2}(f)), 
with the true pattern in Fig.~\ref{fig:simu2}(d), in which 
we averaged the absolute values of the real and imaginary parts 
of the true coefficients across the source points in the ROI. 
Again, STFT-R yielded a sparse activation pattern similar to the truth, 
where as MNE-R yielded a more dispersed pattern. \\

\newcommand{\nonzero}{1}
\begin{figure}[htbp]
\centering
\begin{tabular}{ccc}
\includegraphics[width = 0.33\textwidth]{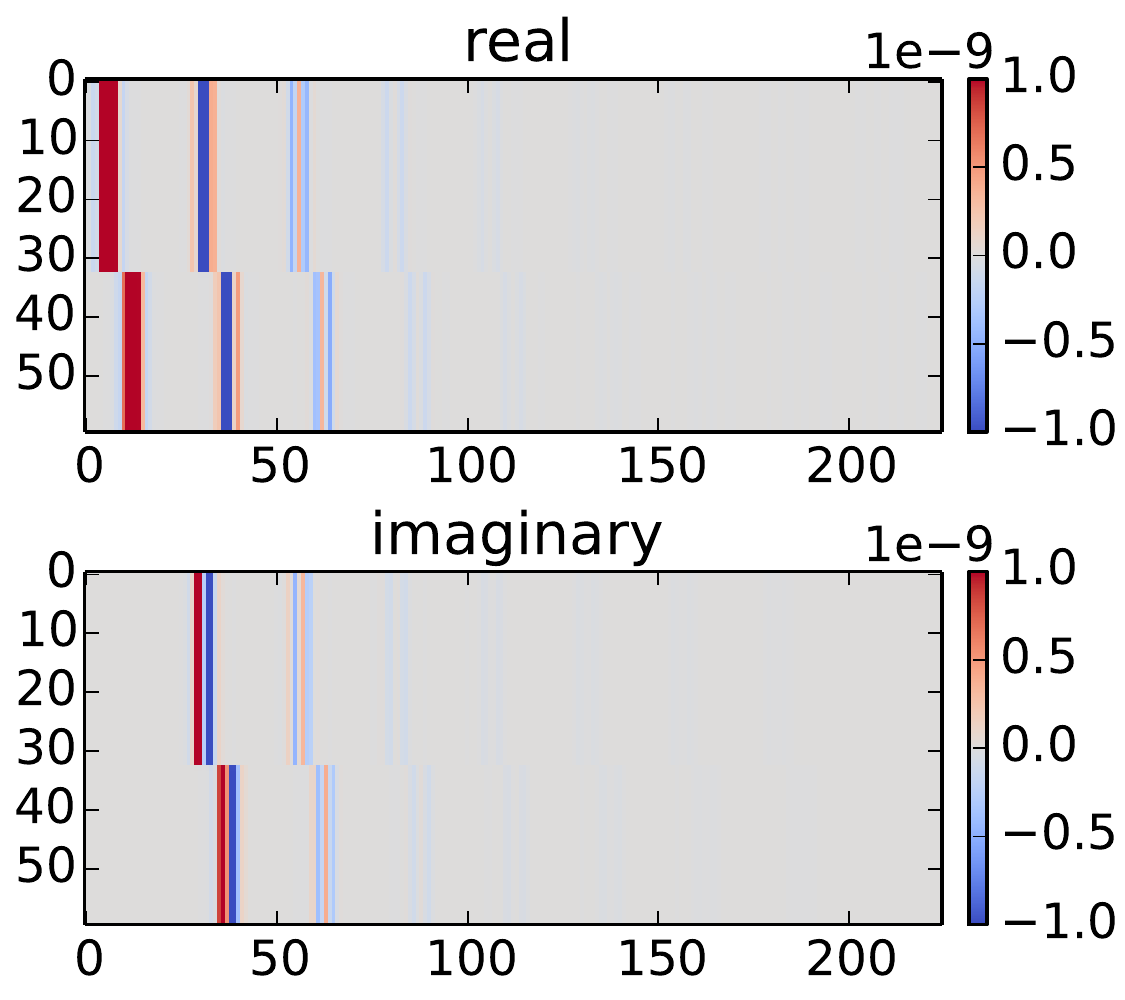} &
\includegraphics[width = 0.33\textwidth]{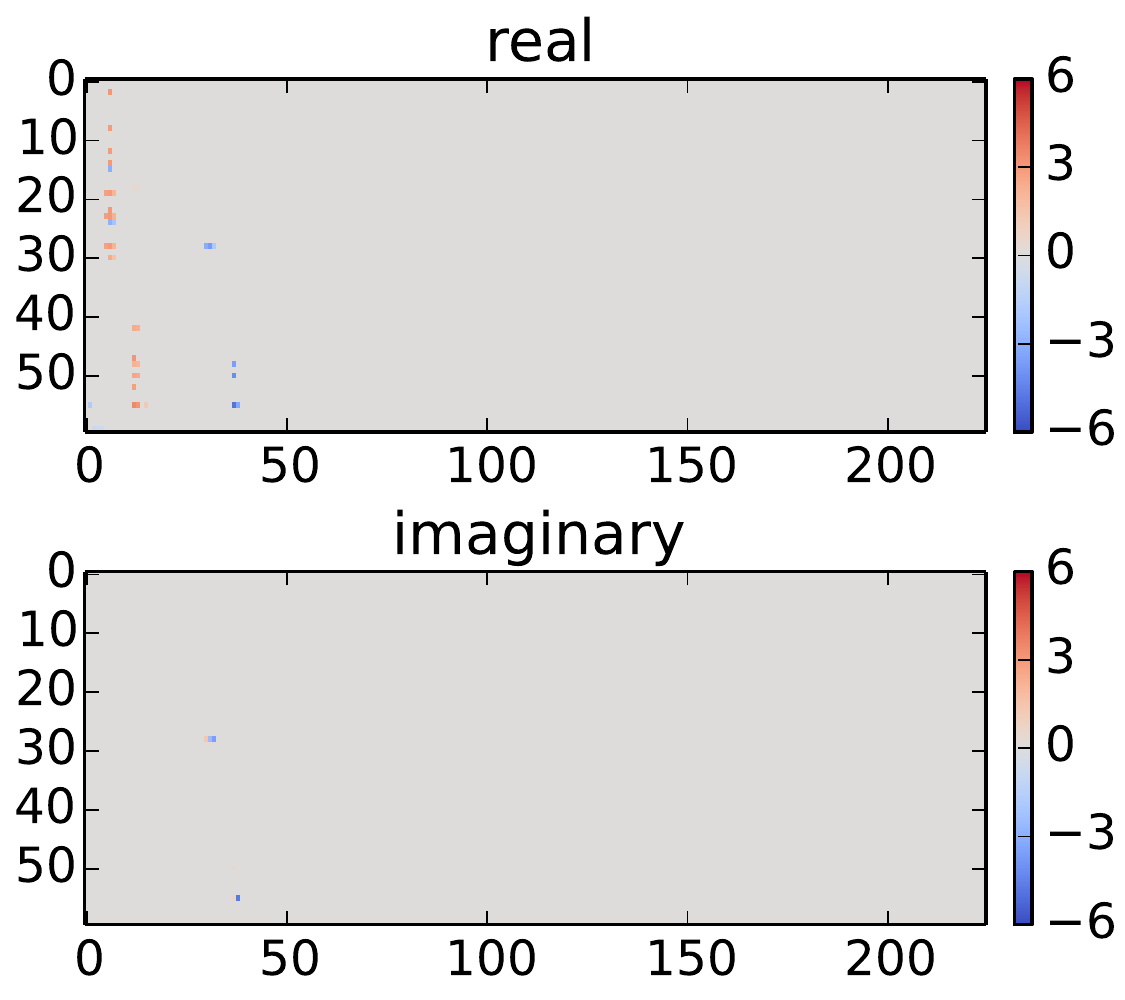} &
\includegraphics[width = 0.33\textwidth]{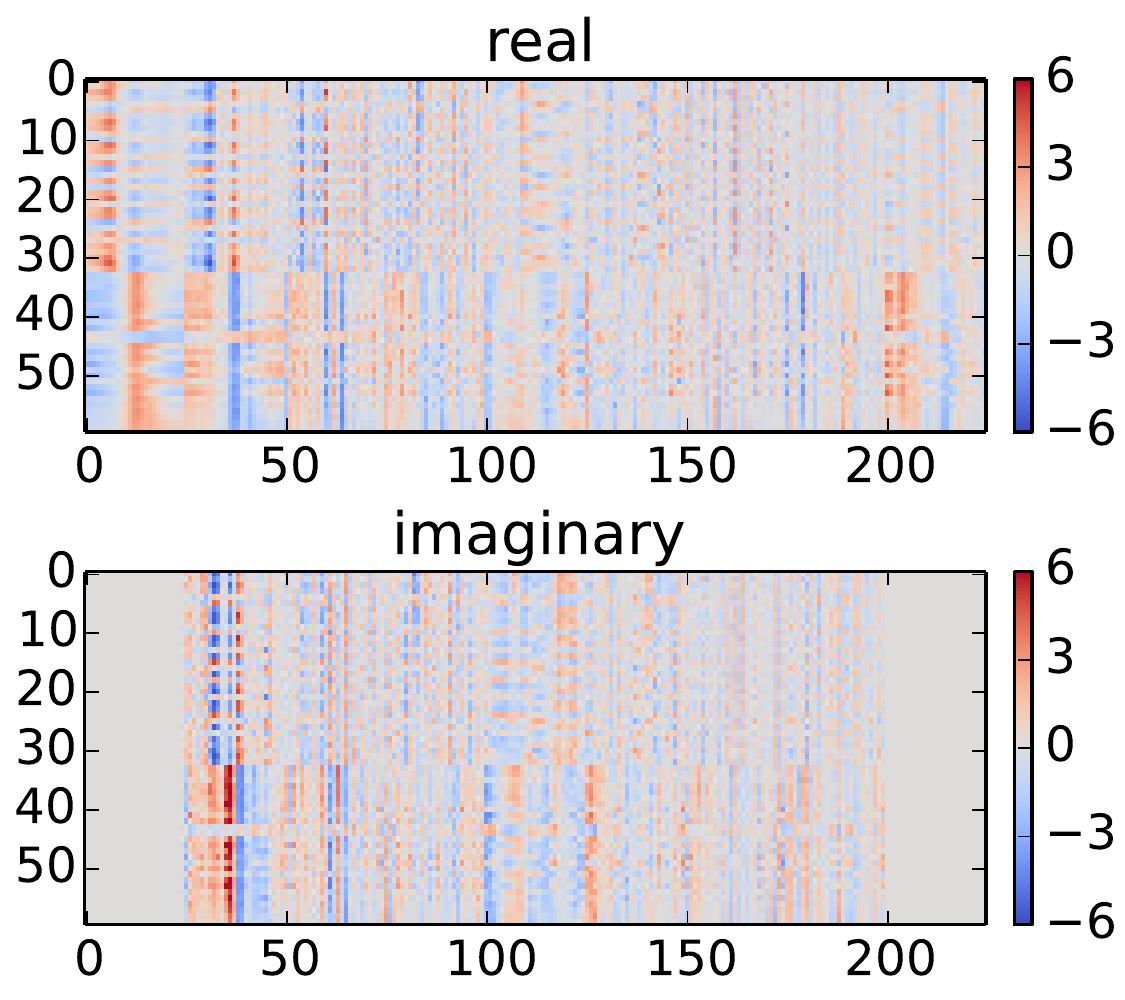} \\
(a) Truth & 
(b) STFT-R &
(c) MNE-R\\
\includegraphics[width = 0.33\textwidth]{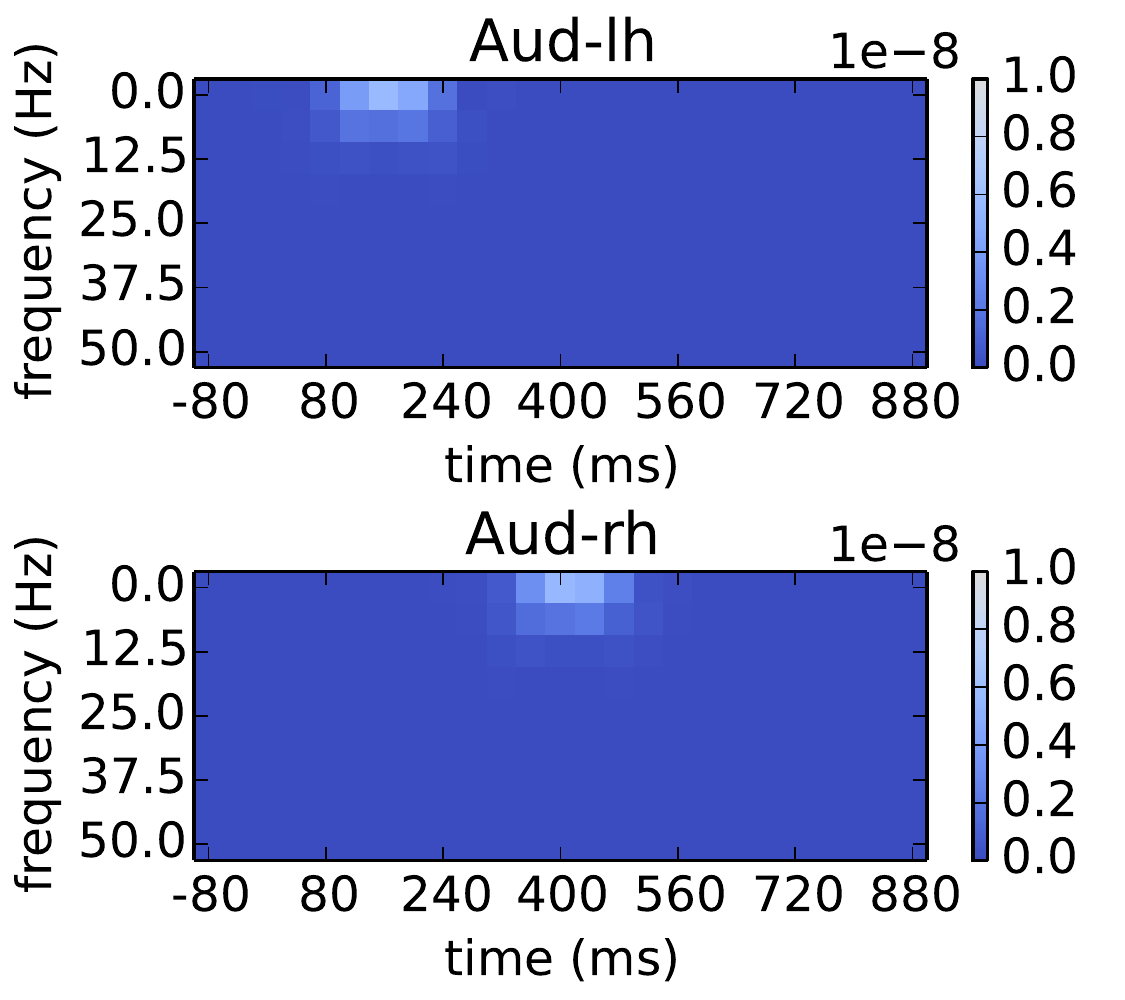} &
\includegraphics[width = 0.33\textwidth]{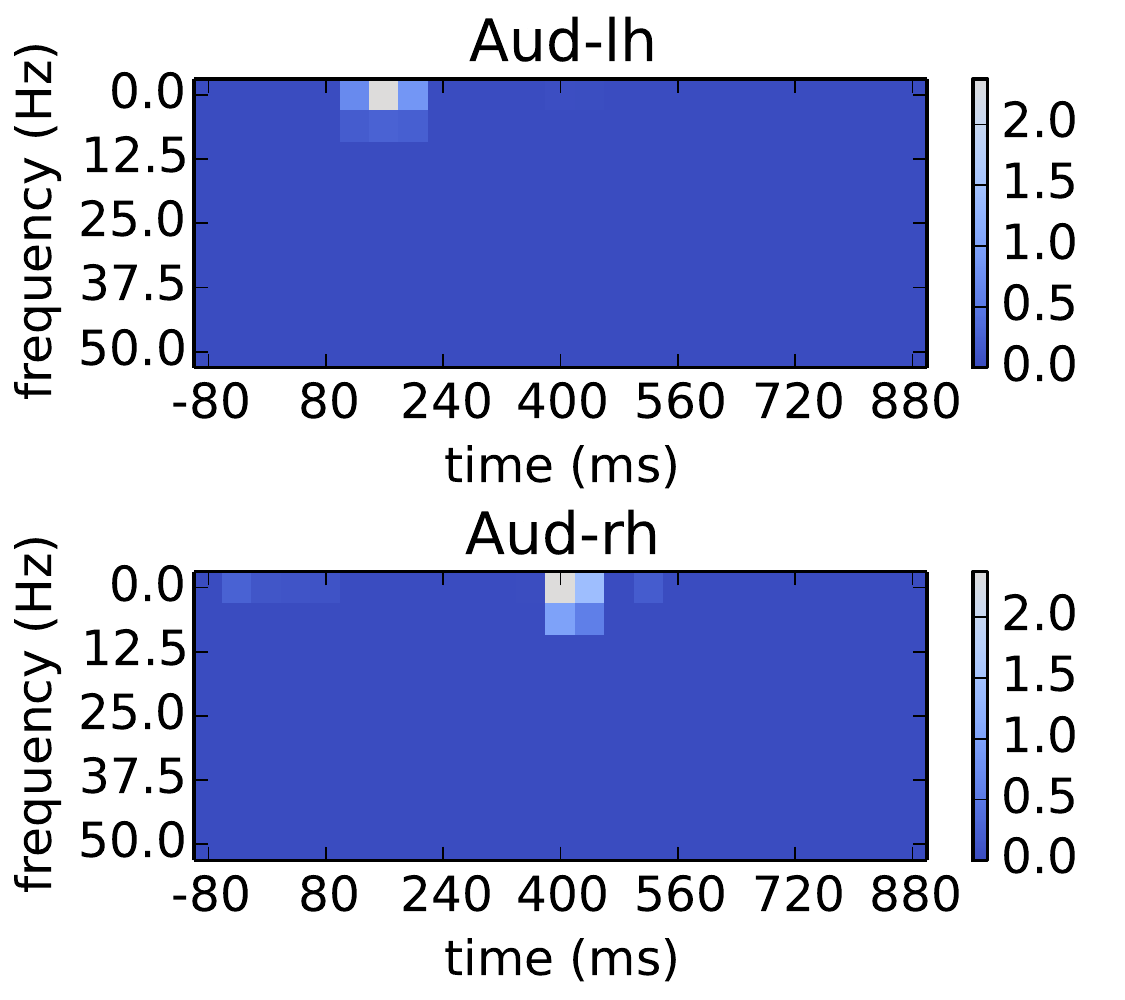} &
\includegraphics[width = 0.33\textwidth]{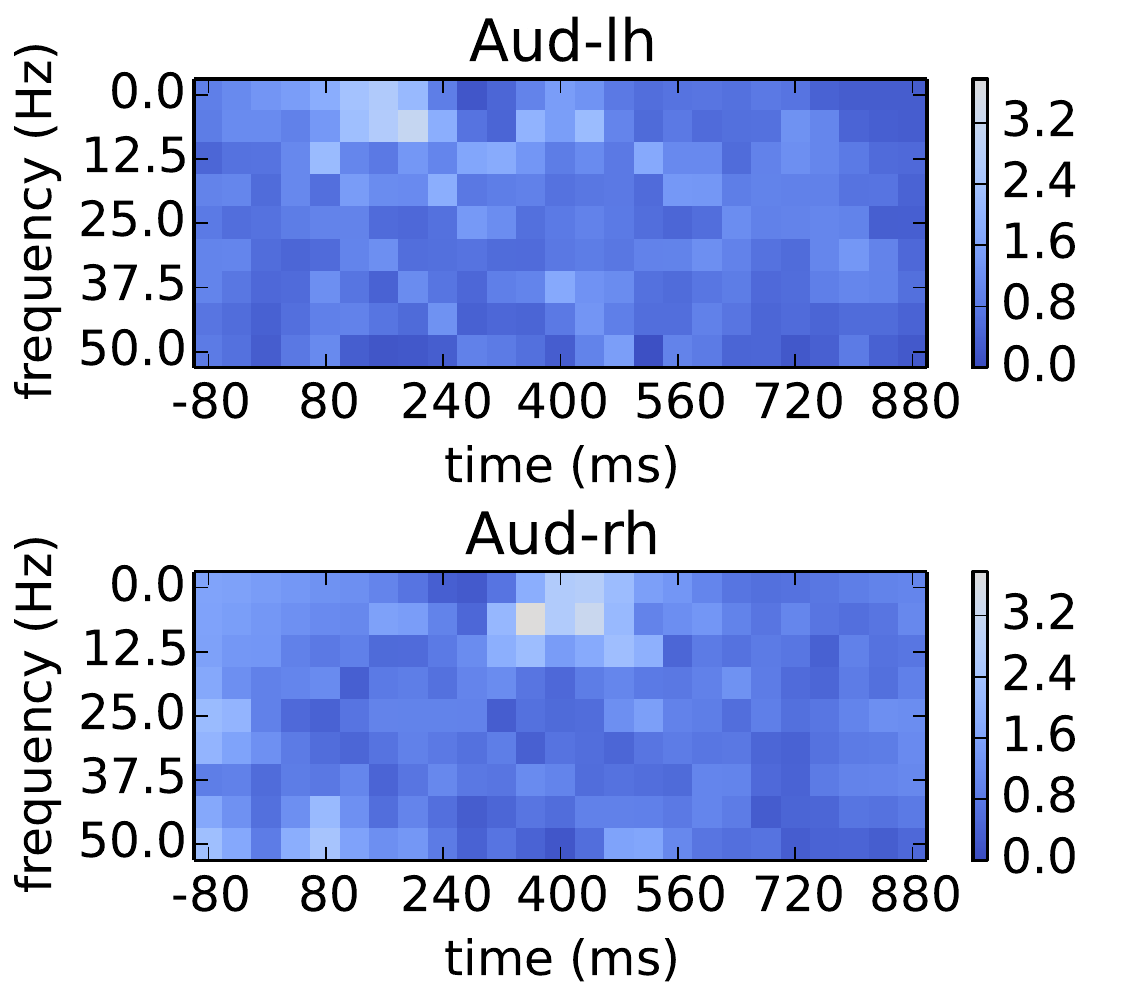} \\
(d) Truth &
(e) STFT-R &
(f) MNE-R \\
\end{tabular}
\caption{Simulation results: inference of regression. \textit{SNR} = \snr, \textit{noise level} = \nratio. 
(a), The true slope coefficients of the regression. 
The vertical axis corresponds to the indices of source points. 
Source points from the two ROIs are concatenated. 
The horizontal axis corresponds to the indices of frequency $\times$ time components, 
where 0-24 are 25 time points in 0 Hz,  25-49 in 6.25 Hz, etc.  
The upper and lower plots show the real and imaginary parts of the complex coefficients. 
(b) and (c), The T-statistics for each STFT components, by STFT-R (b) and by MNE-R(c). 
(d), Averaged absolute values of the real and imaginary parts of the true slope coefficients across source points in each ROI.
(e) and (f), \textit{Averaged absolute T} for each STFT component in the two ROIs by STFT-R (e) and MNE-R (f). }
\label{fig:simu2}
\end{figure}

\indent \textbf{Human face-learning experiment} 
We applied STFT-R and MNE-R 
on a subset of data from a face-learning study \cite{Xu_thesis}, 
where participants learned to distinguish two categories of computer-generated faces. 
In each trial, a participant was shown a face, then reported 
whether it was Category 0 or 1, and got feedback. 
In about 500 trials, participants' behavioural accuracy rates increased from 
chance to at least $70\%$. 
Fig.~\ref{fig:fl1}(a) shows the smoothed 
behavioral accuracy of one participant for Category 0, 
where the smoothing was done by a logistic regression on
Legendre polynomial basis functions of degree 5.  
We used face-selective ROIs pre-defined in an independent dataset, 
and applied STFT-R and MNE-R to regress on the smoothed accuracy rates.
Considering that the participants might have different learning rates for different categories, we analyzed trials with each category separately. 
Again, it was a simple linear regression with only one covariate,
where we fit a slope and an intercept for each STFT component, 
and we were mainly interested in the slope regression coefficients, 
which reflected how neural signals correlated with learning.  
We preprocessed the sensor data using MNE-python and re-sampled the data at 100~Hz. 
STFT was computed in a time window of 160~ms, at a step size $\tau_0 = $ 40~ms. 
When applying STFT, we set the weights of $\alpha$ 
for the ROI groups to zero, 
and used equal weights for other non-ROI groups, 
which summed to 1. 
All of the tuning parameters in both methods, including $\alpha, \beta$ and $\gamma$, were selected via cross-validation. 
We used $B= 20$ in bootstrapping. \\
\newcommand{\ROI}{IOG_R-rh}
\newcommand{\width}{0.32}
\newcommand{\subj}{s1}
\begin{figure}[h]
\centering
\begin{tabular}{cc}
\includegraphics[width =  \width \textwidth]{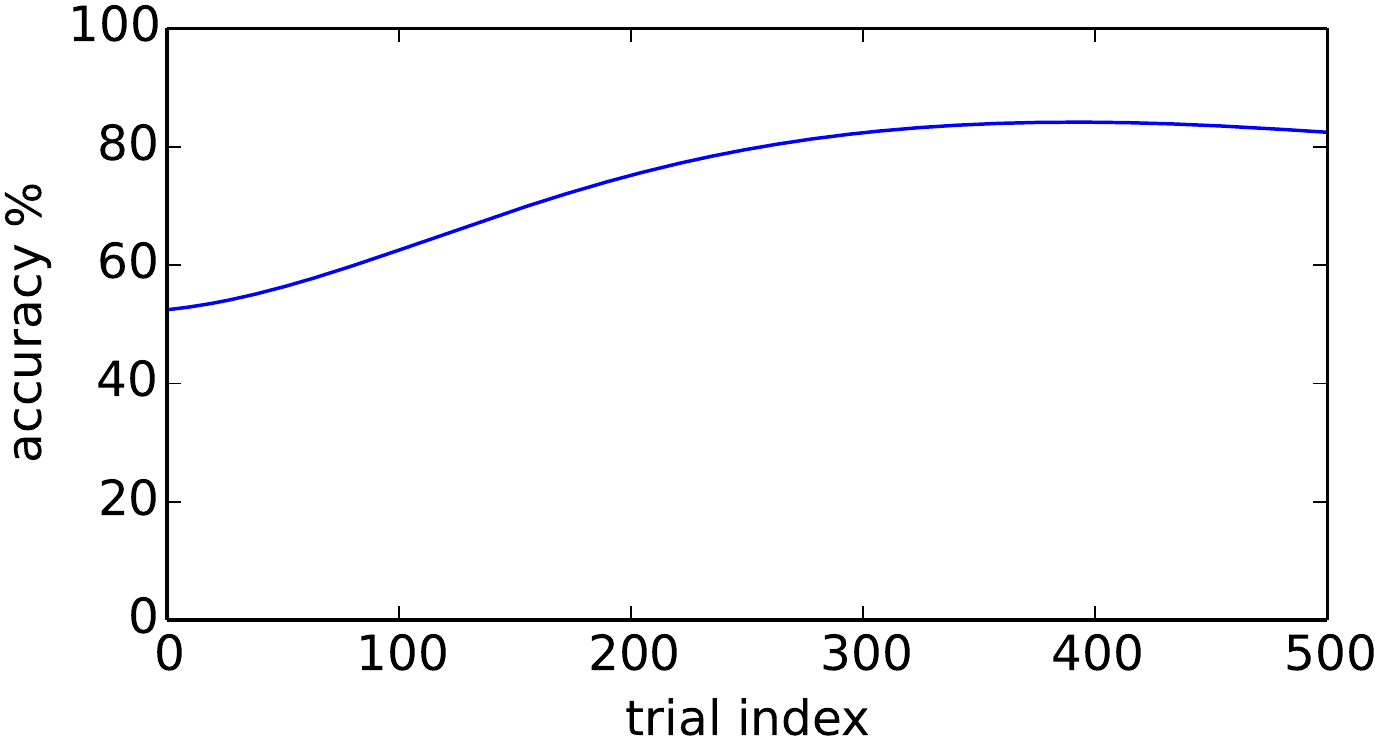} 
&
\\
(a) Behavioral accuracy &\\
\includegraphics[width = \width \textwidth]{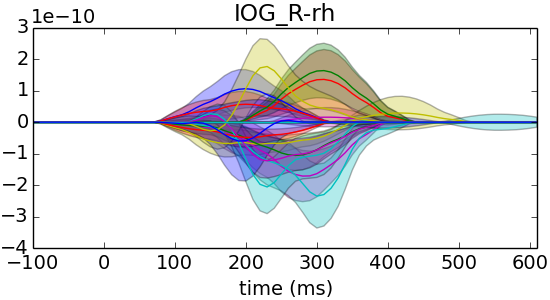}  
&
\includegraphics[width = \width \textwidth]{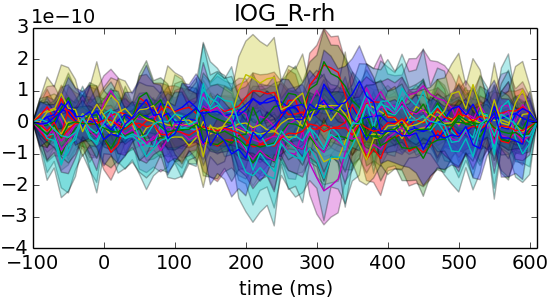} \\
 (b) STFT-R slope time series & (c) MNE-R slope time series \\
\includegraphics[width = \width \textwidth]{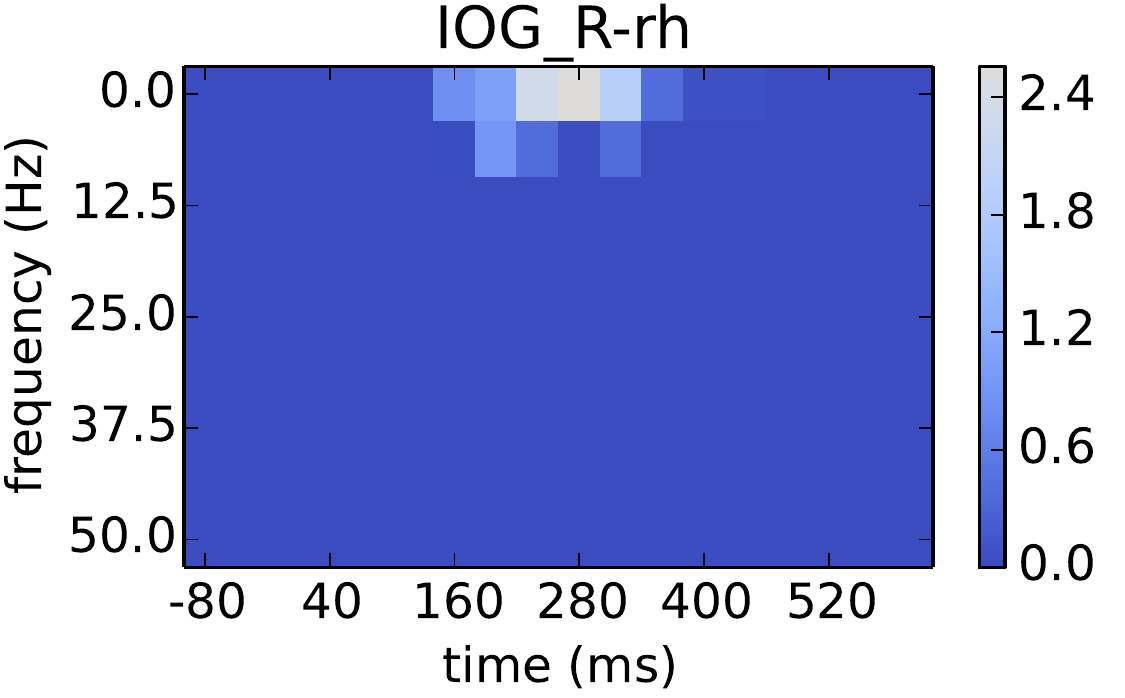}  &
\includegraphics[width = \width \textwidth]{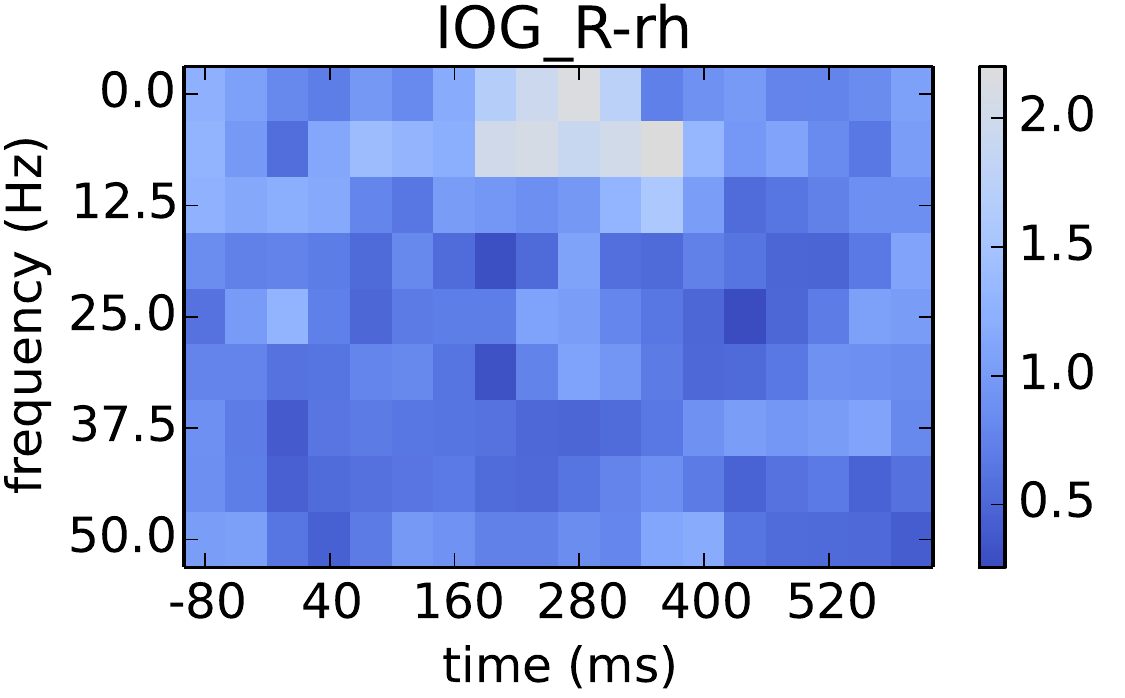}  \\
(d) STFT-R \textit{averaged absolute T} 
& (e) MNE-R \textit{averaged absolute T}\\
\end{tabular}

\caption{Face-learning experiment results for one participant. 
(a), Smoothed behavioral accuracy for Category 0. 
(b) and (c), Time series reconstructed from the STFT slope coefficients in trials with faces of Category 0 . 
Each curve denotes one source point. 
The shaded bands show 95\% confidence intervals. 
(d,e) \textit{Averaged absolute T}s by STFT-R and MNE-R. 
} \label{fig:fl1}
\end{figure}
\indent We report here results in one of the face selective regions, the right inferior occipital gyrus (labelled as \texttt{IOG\_R-rh}), 
for one participant and one face category.
This area is part of the ``occipital face area'' reported in the literature \cite{Pitcher11}. 
Since both STFT and the regression on the
covariates are linear, we inversely transformed the slope coefficients
of the STFT components to a time series for each source point,
(denoted by ``slope time series''), which showed the slope coefficient
in the time domain (Fig.~\ref{fig:fl1}(b) and (c)).  
We observed that
STFT-R produced smooth slope time series due to the sparse STFT
representation (Fig.~\ref{fig:fl1}(b)), whereas MNE-R produced more
noisy time series (Fig.~\ref{fig:fl1}(c)).  
We also show the previously defined \textit{averaged absolute T}s in the ROI, 
produced by STFT-R (Fig.~\ref{fig:fl1}(d)) and MNE-R (Fig.~\ref{fig:fl1}(e))
Compared with the dispersed pattern by
MNE-R, STFT-R produced a more sparse pattern localized near $200 \sim 300$~ms.  
We speculate that this pattern corresponds to the N250 component near 250~ms,
which is related to familiarity of faces \cite{tanaka2006activation}. \\

\section{Discussion} 
To estimate learning effects in MEG, we introduced a source
localization model (STFT-R), in which we embedded regression of STFT
components of source signals on covariates, and exploited a hierarchical
$L_{21}$ penalty to induce structured sparsity and emphasize regions
of interest.  
We derived the FISTA algorithm and an active-set strategy to solve STFT-R. 
In reconstructing the ROI source signals from simulated data, 
STFT-R achieved smaller errors than a two-step method using MNE, 
and in a human learning experiment, 
STFT-R yielded more sparse and thus more interpretable results 
in identifying what time-frequency components of
the ROI signals were correlated with learning. 
In future work, the STFT-R framework can also be used to regress MEG signals on
high-dimensional features of stimuli,
 where the sparsity-inducing property will be able to select important features 
 relevant to complex cognitive processes.\\
\indent
One limitation of STFT-R is its sparse representation of the non-ROI source points.
In our simulation, all of the source points outside the four regions had zero signals, 
and it was reasonable to represent the two irrelevant regions 
as sparse source points. 
However, further simulations are needed to test how well STFT-R behaves 
when the true signals of the non-ROI source points are more dispersed.
It is also interesting to develop a one-step regression model based on 
Bayesian source localization methods \cite{henson2011parametric,mattout2006meg}, 
where we can relax the hard sparse constraints 
but still regularize the problem according to prior knowledge. \\

\subsection*{Appendix 1} \label{Append:STFT}
\textbf{Short-time Fourier transform (STFT)}
Our approach builds on the STFT implemented by Gramfort et al. in \cite{Gramfort13}. 
Given a time series $ \bm U = \{U(t), t = 1,\cdots,T\}$, a time step $\tau_0$ and a window size $T_0$, we define the STFT as
\begin{equation}
\Phi(\{U(t)\},\tau,\omega_h) = \sum_{t=1}^T U(t) K(t-\tau) e^{(-i \omega_h)}
\end{equation}
for $\omega_h = 2\pi h/T_0, h = 0,1,\cdots, T_0/2$
and  $\tau = \tau_0, 2\tau_0, \cdots n_0 \tau_0 $, 
where $K(t-\tau)$ is a window function centered at $\tau$, and 
 $n_0 = T/\tau_0$.  
We concatenate STFT components at different time points and frequencies into a single vector in $ \bm  V \in \mathbb{C}^{s}$, where $s = (T_0/2+1) \times n_0$.  
Following notations in  \cite{Gramfort13}, we also call the $K(t-\tau) e^{(-i \omega_h)}$ terms STFT dictionary functions, and use a matrix's Hermitian transpose $\bm {\Phi^H}$ to denote them, i.e. $ (\bm {U}^T)_{1\times T} = ({\bm V}^T)_{1\times s} (\bm {\Phi ^H})_{s \times T}$. \\
\subsection*{Appendix 2} \label{Append:KKT}
\textbf{ The Karush-Kuhn-Tucker conditions }
Here we derive the Karush-Kuhn-Tucker (KKT) conditions for the hierarchical $L_{21}$ problem. 
Since the term $ f(\bm z) = \frac{1}{2} \sum_{r=1}^q  ||\bm{M} ^{(r)} - \bm G( \sum_{k=1}^p X_k^{(r)} \bm{Z}_k ) \bm{\Phi^H}||_F^2$ is essentially a sum of squared error of a linear problem, 
we can re-write it as $ f(\bm z) = \frac{1}{2} || \bm b - \bm A \bm z ||^2$, where $\bm z$ again is a vector concatenated by entries in $\bm Z$, $\bm b$ is a vector concatenated by $\bm M^{(1)}, \cdots, \bm M^{(q)}$, and $ \bm A$ is a linear operator, such that $\bm A \bm z$ is the concatenated $\bm G( \sum_{k=1}^p X_k^{(r)} \bm{Z}_k ) \bm{\Phi^H}, r = 1,\cdots, q$. 
Note that although $\bm z$ is a complex vector, we can further reduce the problem into a real-valued problem by rearranging the real and imaginary parts of $\bm z$ and $ \bm A$. 
Here for simplicity, we only derive the KKT conditions for the real case. 
Again we use $\{g_1, \cdots, g_h, \cdots, g_N \}$ to denote our ordered hierarchical group set, and $\lambda_h$ to denote the corresponding penalty for group $g_h$. 
We also define diagonal matrices  $\bm{D}_h$ such that
\begin{align*} 
 \bm{D}_h(l,l) = \left\{
   \begin{array}{l l} 
    1 &\text{ if } l\in g_h  \\
  0  & \text{ otherwise } 
  \end{array} \right. \forall h 
\end{align*}
therefore, the non-zero elements of $\bm{D}_h \bm{z}$ is equal to $\bm{z} |_{g_h}$. 
With the simplified notation, we re-cast the original problem into a standard formulation: 
\begin{align}
\min_{\bm{z}}  (\frac{1}{2}\|\bm b - \bm A \bm z \|^2_2 + \sum_h \lambda_h \|\bm{D}_h \bm{z} \|_2) \label{eq: tree_group_lasso}
\end{align}
To better describe the KKT conditions, we introduce some auxiliary variables, $\bm{u} = \bm A \bm z, \bm{v}_h = \bm{D}_h \bm z$. 
Then (\ref{eq: tree_group_lasso}) is equivalent to 
\begin{align*}
\min_{\bm{z}, \bm{u} , \bm{v}_h}   & (\frac{1}{2}\|\bm b - \bm u \|^2_2 + \sum_h \lambda_h \| \bm{v}_h \|_2) \\
\text{such that } &  \bm{u} = \bm A \bm z, \quad 
                   \bm {v}_h= \bm{D}_h \bm{z}, \forall h    
\end{align*}
The corresponding Lagrange function is 
\begin{align*}
L(\bm{z}, \bm{u} , \bm{v}_h,\bm{\mu},\bm{\xi}_h  ) =   \frac{1}{2}\|\bm b - \bm u \|^2_2 + \sum_h \lambda_h\| \bm{v}_h \|_2  +   \bm{\mu}^T ( \bm A \bm z - \bm{u}) +  \sum_{h} \bm{\xi}_h^T ( \bm{D}_h \bm{z} - \bm {v}_h )   
\end{align*}
where $\bm{\mu}$ and $\bm{\xi}_h$'s are Lagrange multipliers. 
At the optimum, the following KKT conditions hold
\begin{align}
\frac{\partial{L}}{\partial{\bm{u}}} & =  \bm{u} - \bm{b} - \bm{\mu} = 0  \label{eq: KKT_mu}\\
\frac{\partial{L}}{\partial{\bm{z}}} & =   \bm{A}^T \bm{\mu} + \sum_h \bm{D}_h \bm{\xi}_h  = 0 \label{eq: KKT_mu_xi}\\
\frac{\partial{L}}{\partial{\bm{v}_h}} & = \lambda_h  \partial{ \|\bm{v}_h \|_2} - \bm{\xi}_h \ni 0,  \forall h \label{eq:KKT_xi}
\end{align}
where $\partial{ \| \cdot \|_2}$ is the subgradient of the $L_2$ norm. 
From  (\ref{eq: KKT_mu}) we have $ \bm{\mu} = \bm{u} - \bm{b}$, then (\ref{eq: KKT_mu_xi}) becomes $ \bm{A}^T (\bm{u} - \bm{b}) + \sum_h \bm{D}_h \bm{\xi}_h  = 0 $.  
Plugging $\bm{u}= \bm{Az}$ in, we can see that  the first term 
$  \bm{A}^T ( \bm{u} - \bm{b}) = \bm{A}^T (\bm{A} \bm{z} - \bm{b})$ is the gradient of $ f(\bm z) = \frac{1}{2} \| \bm b - \bm{A} \bm{z}\|_2^2$. 
For a solution $\bm {z}_0$, once we plug in $ \bm{v}_h = \bm{D}_h \bm{z}_0 $, the KKT conditions become
\begin{align}
&\nabla f(\bm{z})_{\bm z = \bm{z}_0} + \sum_h \bm{D}_h \bm{\xi}_h  = 0 \label{eq:KKT1}\\
&\lambda_h  \partial{ \| \bm{D}_h \bm{z}_0 \|_2} - \bm{\xi}_h \ni 0, \forall h \label{eq:KKT2}
\end{align}
In (\ref{eq:KKT2}), we have the following according to the definition of subgradients
\begin{align*}
&\bm{\xi}_h =  \lambda_h \frac{ \bm{D}_h \bm{z}_0 } {\| \bm{D}_h \bm{z}_0 \|_2} \text{  if  } \| \bm{D}_h \bm{z}_0 \|_2 > 0 \\
&\|\bm{\xi}_h\|_2 \le  \lambda_h  \text{  if  } \| \bm{D}_h \bm{z}_0 \|_2 = 0
\end{align*}
Therefore we can determine whether (\ref{eq:KKT1}) and (\ref{eq:KKT2}) hold by solving the following problem. 
\begin{align*}
\min_{\bm{\xi}_h}  & \frac{1}{2} \|\nabla f(\bm{z})_{\bm z = \bm{z}_0} + \sum_h \bm{D}_h \bm{\xi}_h\|_2^2 \\
\text{subject to } & \bm{\xi}_h =  \lambda_h \frac{ \bm{D}_h \bm{z}_0 } {\| \bm{D}_h \bm{z}_0 \|_2} \text{  if  } \| \bm{D}_h \bm{z}_0 \|_2 > 0 \\
&\|\bm{\xi}_h\|_2 \le  \lambda_h  \text{  if  } \| \bm{D}_h \bm{z}_0 \|_2 = 0
\end{align*}
which is a standard group lasso problem with no overlap. 
We can use coordinate-descent to solve it. 
We define $\frac{1}{2} \|\nabla f(\bm{z})_{\bm z = \bm{z}_0} + \sum_h \bm{D}_h \bm{\xi}_h\|_2^2$ at the optimum as a measure of violation of the KKT conditions.\\ 
\indent
Let $f_{J}$ be the function $f$ constrained on a set $J$. 
Because the gradient of $f$ is linear, 
if $\bm{z_0}$ only has non-zero entries in $J$, then
the entries of $\nabla f( \bm{z})$ in $J$ are equal to $\nabla f_{J} (\bm{z} |_J)$ at $\bm z = \bm{z}_0$. 
In addition, $\bm{\xi}_h$'s are separate for each group. 
Therefore if $\bm{z}_0$ is an optimal solution to the problem constrained on $J$,  then the KKT conditions are already met for entries in $J$ (i.e. $ \left(\nabla f(\bm{z})_{\bm z = \bm{z}_0} + \sum_h \bm{D}_h \bm{\xi}_h\right)|_{J}  = 0$);
for $g_h \not\subset J$, 
we use ( $\frac{1}{2}\|\left(\nabla f(\bm{z})_{\bm z = \bm{z}_0} + \sum_h \bm{D}_h \bm{\xi}_h\right)|_{g_h} \|^2$) at the optimum as a measurement of how much the elements in group $g_h$ violate the KKT conditions, which is a criterion when we greedily add groups (see Algorithm~\ref{alg:active_set}).

\section*{Acknowledgements}
This work was funded by the Multi-Modal Neuroimaging Training Program (MNTP)
fellowship from the NIH ( 5R90DA023420-08,5R90DA023420-09) and Richard King Mellon
Foundation.
We also thank Yang Xu and the MNE-python user group for their help. 
\bibliography{GroupLassoSTFT_ref.bib}

\begin{thebibliography}{}

\bibitem[Bach et~al., 2011]{Bach11}
Bach, F., Jenatton, R., Mairal, J., and Obozinski, G. (2011).
\newblock Optimization with sparsity-inducing penalties.
\newblock {\em CoRR}, abs/1108.0775.

\bibitem[Beck and Teboulle, 2009]{fista}
Beck, A. and Teboulle, M. (2009).
\newblock A fast iterative shrinkage-thresholding algorithm for linear inverse
  problems.
\newblock {\em SIAM Journal on Imaging Sciences}, 2(1):183--202.

\bibitem[Dale et~al., 2000]{Dale00}
Dale, A.~M., Liu, A.~K., Fischl, B.~R., Buckner, R.~L., Belliveau, J.~W.,
  Lewine, J.~D., and Halgren, E. (2000).
\newblock Dynamic statistical parametric mapping: combining fmri and meg for
  high-resolution imaging of cortical activity.
\newblock {\em Neuron}, 26(1):55--67.

\bibitem[Galka et~al., 2004]{Galka04}
Galka, A., Ozaki, O. Y.~T., Biscay, R., and Valdes-Sosa, P. (2004).
\newblock A solution to the dynamical inverse problem of eeg generation using
  spatiotemporal kalman filtering.
\newblock {\em NeuroImage}, 23:435--453.

\bibitem[Gauthier et~al., 2000]{Gauthier00}
Gauthier, I., Tarr, M.~J., Moylan, J., Skudlarski, P., Gore, J.~C., and
  Anderson, A.~W. (2000).
\newblock The fusiform “face area” is part of a network that processes
  faces at the individual level.
\newblock {\em Journal of cognitive neuroscience}, 12(3):495--504.

\bibitem[Gramfort et~al., 2014]{mne-python}
Gramfort, A., Luessi, M., Larson, E., Engemann, D.~A., Strohmeier, D.,
  Brodbeck, C., Parkkonen, L., and Hämäläinen, M.~S. (2014).
\newblock Mne software for processing meg and eeg data.
\newblock {\em NeuroImage}, 86(0):446 -- 460.

\bibitem[Gramfort et~al., 2013]{Gramfort13}
Gramfort, A., Strohmeier, D., Haueisen, J., Hamalainen, M., and Kowalski, M.
  (2013).
\newblock Time-frequency mixed-norm estimates: Sparse m/eeg imaging with
  non-stationary source activations.
\newblock {\em NeuroImage}, 70(0):410 -- 422.

\bibitem[Hamalainen et~al., 1993]{Hamalainen93}
Hamalainen, M., Hari, R., Ilmoniemi, R.~J., Knuutila, J., and Lounasmaa, O.~V.
  (1993).
\newblock Magnetoencephalography--theory, instrumentation, to noninvasive
  studies of the working human brain.
\newblock {\em Reviews of Modern Physics}, 65:414--487.

\bibitem[Hamalainen and Ilmoniemi, 1994]{Hamalainen94}
Hamalainen, M. and Ilmoniemi, R. (1994).
\newblock Interpreting magnetic fields of the brain: minimum norm estimates.
\newblock {\em Med. Biol. Eng. Comput.}, 32:35--42.

\bibitem[Henson et~al., 2011]{henson2011parametric}
Henson, R.~N., Wakeman, D.~G., Litvak, V., and Friston, K.~J. (2011).
\newblock A parametric empirical bayesian framework for the eeg/meg inverse
  problem: generative models for multi-subject and multi-modal integration.
\newblock {\em Frontiers in human neuroscience}, 5.

\bibitem[Jenatton et~al., 2011]{Jenatton11a}
Jenatton, R., Mairal, J., Obozinski, G., and Bach, F. (2011).
\newblock Proximal methods for hierarchical space coding.
\newblock {\em J. Mach. Learn. Res}, 12:2297--2334.

\bibitem[Kanwisher et~al., 1997]{Kanwisher97}
Kanwisher, N., McDermott, J., and Chun, M.~M. (1997).
\newblock The fusiform face area: a module in human extrastriate cortex
  specialized for face perception.
\newblock {\em The Journal of Neuroscience}, 17(11):4302--4311.

\bibitem[Lamus et~al., 2012]{Lamus12}
Lamus, C., Hamalainen, M.~S., Temereanca, S., Brown, E.~N., and Purdon, P.~L.
  (2012).
\newblock A spatiotemporal dynamic distributed solution to the meg inverse
  problem.
\newblock {\em NeuroImage}, 63:894--909.

\bibitem[Mattout et~al., 2006]{mattout2006meg}
Mattout, J., Phillips, C., Penny, W.~D., Rugg, M.~D., and Friston, K.~J.
  (2006).
\newblock Meg source localization under multiple constraints: an extended
  bayesian framework.
\newblock {\em NeuroImage}, 30(3):753--767.

\bibitem[Pascual-Marqui, 2002]{Pascual-Marqui02}
Pascual-Marqui, R. (2002).
\newblock Standardized low resolution brain electromagnetic tomography
  (sloreta): technical details.
\newblock {\em Methods Find. Exp. Clin. Pharmacol.}, 24:5--12.

\bibitem[Pitcher et~al., 2011]{Pitcher11}
Pitcher, D., Walsh, V., and Duchaine, B. (2011).
\newblock The role of the occipital face area in the cortical face perception
  network.
\newblock {\em Experimental Brain Research}, 209(4):481--493.

\bibitem[Stine, 1985]{Stine85}
Stine, R.~A. (1985).
\newblock Bootstrp prediction intervals for regression.
\newblock {\em Journal of the American Statistical Association}, 80:1026--1031.

\bibitem[Tanaka et~al., 2006]{tanaka2006activation}
Tanaka, J.~W., Curran, T., Porterfield, A.~L., and Collins, D. (2006).
\newblock Activation of preexisting and acquired face representations: the n250
  event-related potential as an index of face familiarity.
\newblock {\em Journal of Cognitive Neuroscience}, 18(9):1488--1497.

\bibitem[Xu, 2013]{Xu_thesis}
Xu, Y. (2013).
\newblock Cortical spatiotemporal plasticity in visual category learning
  (doctoral dissertation).

\end{thebibliography}
\bibliographystyle{apalike}

\end{document}